\documentclass{article}
\usepackage{spconf,amsmath,graphicx}

\usepackage{hyperref}
\usepackage[utf8]{inputenc}
\usepackage{tabularx}
\usepackage{setspace}
\usepackage{graphicx}
\usepackage{booktabs}
\usepackage{amsmath}
\usepackage{color}
\usepackage{multirow}

\usepackage{url,cite} 
\usepackage{caption}
\usepackage{verbatim} % for block comment
\usepackage{amssymb}
\usepackage{algorithm}
\usepackage{algorithmic}
\usepackage{tikz}
\usepackage{footnote}
\usepackage{url}

\title{Learning to match transient sound events using attentional similarity for few-shot sound recognition}
\name{Szu-Yu Chou$^{1,2}$, Kai-Hsiang Cheng$^2$,
Jyh-Shing Roger Jang$^1$, 
Yi-Hsuan Yang$^2$}
\address{$^1$Graduate Institute of Networking and Multimedia, National Taiwan University, Taipei, Taiwan \\
$^2$Research Center for Information Technology Innovation, Academia Sinica, Taipei, Taiwan}

\begin{document}

%\ninept
%
\maketitle
\begin{abstract}
In this paper, we introduce a novel attentional similarity module for the problem of few-shot sound recognition. Given a few examples of an unseen sound event, a classifier must be quickly adapted to recognize the new sound event without much fine-tuning. The proposed attentional similarity module can be plugged into any metric-based learning method for few-shot learning, allowing the resulting model to especially match related short sound events. Extensive experiments on two datasets show that the proposed module consistently improves the performance of five different metric-based learning methods for few-shot sound recognition. The relative improvement ranges from +4.1\% to +7.7\% for 5-shot 5-way accuracy for the ESC-50 dataset, and from +2.1\% to +6.5\% for \textit{noise}ESC-50. Qualitative results demonstrate that our method contributes in particular to the recognition of transient sound events.

\end{abstract}
\section{Introduction}
\label{sec:intro}

Understanding the surrounding environment through sounds has been considered as a major component in many daily applications, such as surveillance, smart city, and smart cars \cite{Mesaros2016TUTDF,DCASE17}. Recent years have witnessed great progress in sound event detection and classification using deep learning techniques \cite{Liu16MM,Hershey17icassp,Vu2017,Lee2017a,su17icassp}. However, most prior arts rely on standard supervised learning algorithm and may not perform well for sound events with sparse training examples. 
While such a \emph{few-shot learning} task has been increasingly studied in neighboring fields such as computer vision and natural language processing \cite{matching_network}, little work if any has been done for few-shot sound event recognition, to our best knowledge.

\begin{figure}[t]
\centering
\includegraphics[width=.49\textwidth]{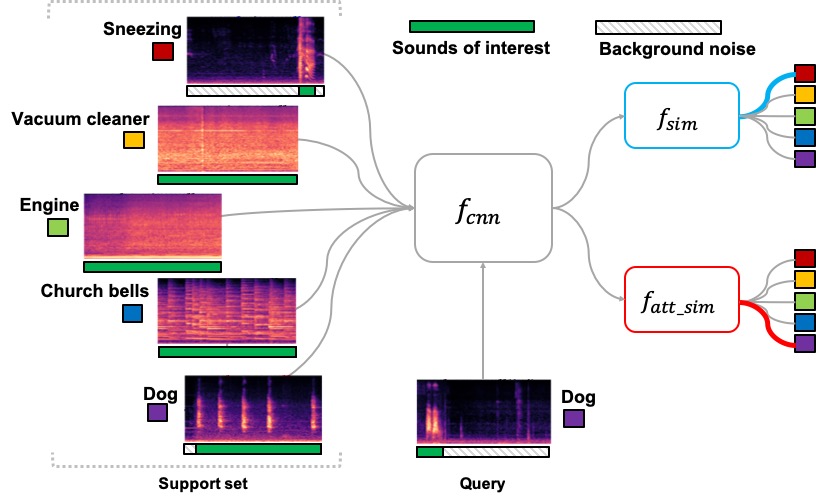}
\caption{Illustration of using the proposed attentional similarity module (marked in red) for few-shot sound recognition. Compared to common similarity $f_{sim}$, the attentional similarity $f_{att\_sim}$ performs better in matching transient (e.g., less than 1 second) sound events. Moreover, the attentional similarity module can be applied to any network for few-shot learning. The squares of different color denote different sound events. All the figures in this paper are best viewed in color.}
\label{fig:sse} 
\end{figure}

Several metric-based learning methods for general few-shot learning have been proposed in the literature \cite{Koch2015SiameseNN,matching_network,sung2018learning,Jake17nips}. The \textit{Matching Network} proposed by Vinyals \emph{et al.} \cite{matching_network} uses the cosine similarity to measure the distance of a learned representation of the labeled set of examples with a giving unlabeled example for classification. Importantly, it proposes an \textit{episode} procedure during training, which samples only a few examples of each class as data points to simulate the few-shot learning scenario. Such a procedure allows the training phase to be close to the test phase in few-shot learning and accordingly improves the model's generalization ability. Snell \emph{et al.} \cite{Jake17nips} follows the episodic procedure and proposes the \textit{Prototypical Network}. They take the average of the learned representation of a few examples for each class as a class-wise representation, and then classifies an input unlabeled datum by calculating the Euclidean distance between the input and the class-wise representations. Sung \emph{et al.} \cite{sung2018learning} proposes the \textit{Relation Network} to learn a non-linear distance metric for measuring the distance between the input unlabeled examples and a few examples from each class.

Most sound event recognition models are trained on datasets with clip-level labels, such as DCASE2016 \cite{Mesaros2016TUTDF}, ESC-50 \cite{piczak2015dataset}, and AudioSet \cite{audioset}. 
Because the clip-level labels do not specify where the corresponding event actually takes place in an audio signal, this training strategy may make a model overlook short or transient sound events \cite{chou18ijcai}. To tackle this issue for few-shot sound recognition, in this paper, we propose a novel \emph{attentional similarity} module to automatically guide the model to pay attention to specific segments of a long audio clip for recognizing relatively short or transient sound events. We show that our attentional similarity module can be learned relying on only clip-level annotation, and that it can be plugged into any existing methods to improve their performance for few-shot sound recognition.

\section{Approach}
\label{sec:approach}

\subsection{Few-shot sound recognition}

\begin{figure}[t]
\centering
\includegraphics[width=.49\textwidth]{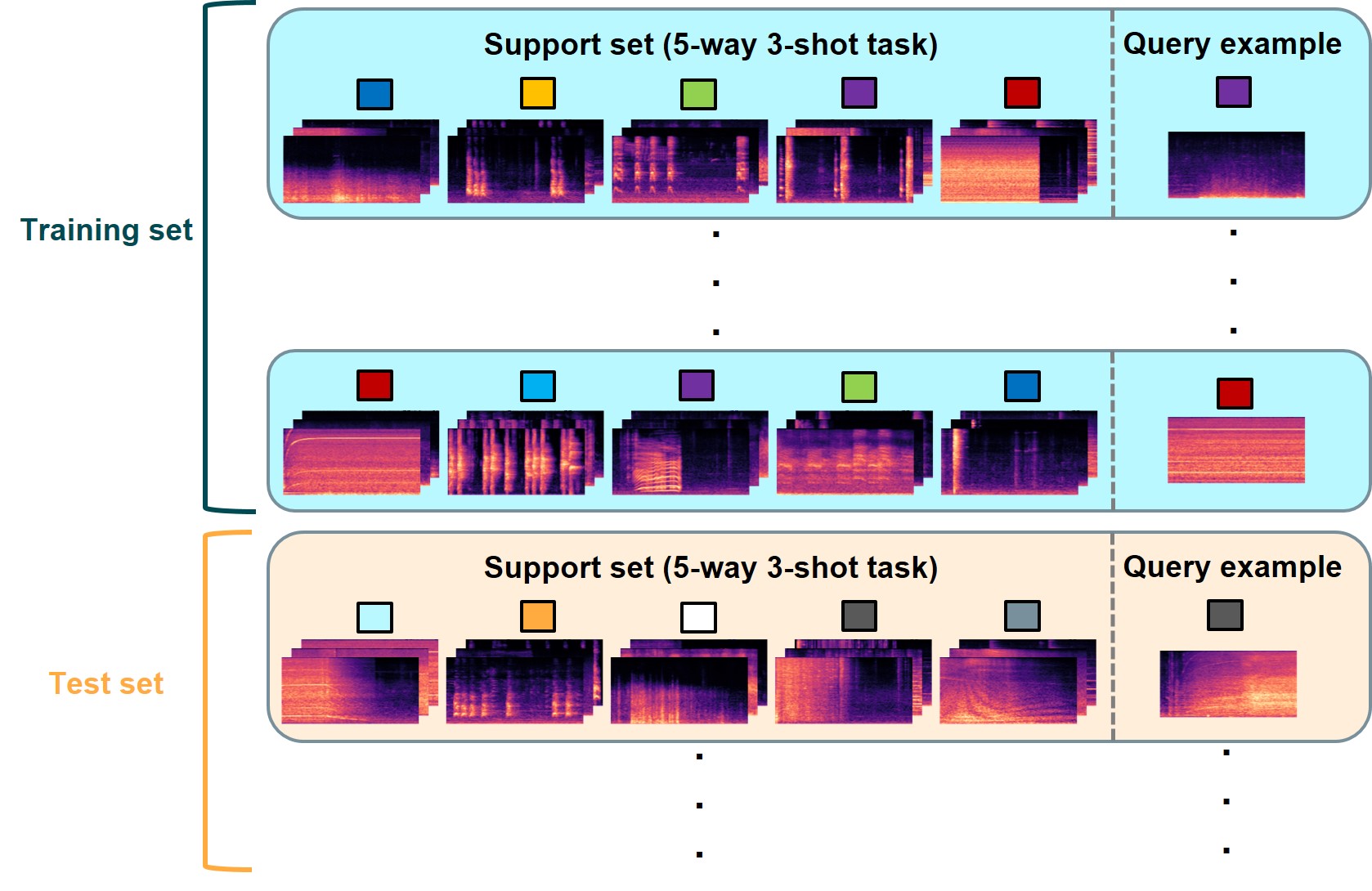}
\caption{The training and test sets for few-shot sound recognition. The classes of test set are not seen during training.}
\label{fig:tr_te} 
\end{figure}

The goal of few-shot sound recognition is to learn a classifier which can quickly accommodate to unseen classes with only a few examples. In training, we are given a training set $\mathcal{D}=\{(\chi,\mathbf{y})\mid \chi \in \mathcal{S}, \mathbf{y} \in C_{train} \}$, where $\mathcal{S}$ is a support set, $y$ denotes the class of the support set and $C_{train}$ is the total number of classes in the training set. Generally, the support set $\mathcal{S}=\{(s_i,...,s_N) \mid s_i \in \{x_1,...,x_{k \times c},x_q\} \}$ consists of small support examples and a query example $x_q$ (see Fig. \ref{fig:tr_te}), where $x$ is the input feature and $N$ is the number of support sets. The support examples are randomly sampled $k$ examples from each of the $c$ classes in the training set $\mathcal{D}$, and the query sample is randomly chosen from the remaining examples of $c$ classes. The task is called \emph{$c$-way $k$-shot learning}. And, $k$ is often a small number from 1 to 5.

In our work, we use the simple yet powerful ConvNet architecture as our feature learning model $f_{cnn}(\cdot)$. The model can be learnt by minimizing the following objective function:
\begin{equation}
\min_{\theta} \sum_{(\chi,\mathbf{y})\in \mathcal{D}}  \mathcal{L}(f_{cnn}(\chi),\mathbf{y})   + \mathcal{R}(\theta)\,,
\label{eq_multitask}
\end{equation}
where $\mathcal{L}$ is a loss function, $\theta$ is the parameters of network and $\mathcal{R}(\theta)$ is a
regularization term for avoiding overfitting.

Similar to the sate-of-the-art algorithms \cite{matching_network,Jake17nips,sung2018learning} for few-shot learning, our loss function is based on the cross entropy:
\begin{equation}
\mathcal{L} = \frac{\textrm{exp}(\sum_{\mathbf{X}_j \in S_y}f_{sim}(\mathbf{X}_q,\mathbf{X}_j)}
{\sum_{j=1}^c \textrm{exp}(\sum_{\mathbf{X}_j \in S_j}f_{sim}(\mathbf{X}_q,\mathbf{X}_j)}  \,,
\label{eq_multitask}
\end{equation}
where $\mathbf{X}$ is the output (a.k.a. a \emph{feature map}) from the last convolutional layer of $f_{cnn}$, $S_y$ denotes the set of inputs labeled with class $y$ and $f_{sim}$ is the similarity function for measuring the distance between two inputs.

\begin{table*}[t]
\centering
%The \textbf{top}-half compares the performance with several non-attention methods. The \textbf{bottom}-half demonstrates the performance by compared prior art attention networks and the proposed network. We also report the number of depth and parameters of the network.
\begin{tabular}{|l|ccc|cccc|}
\hline
\multirow{2}{*}{Model} & Attentional &\multirow{2}{*}{Depth} &\multirow{2}{*}{Param.}  &\multicolumn{2}{c}{5-way Acc} &\multicolumn{2}{c|}{10-way Acc} \\
\cline{5-8}
  &similarity&&  &1-shot &5-shot &1-shot &5-shot  \\
\hline
\textsc{Siamese Network} \cite{Koch2015SiameseNN}     &     &3  &1.18M   &43.5\%  &50.9\% &26.1\% &31.6\%\\
\textsc{Matching Network} \cite{matching_network}    &     &3  &1.18M   &53.7\%  &67.0\% &34.5\% &47.9\%\\
\textsc{Relation Network} \cite{sung2018learning}     &     &7  &4.78M   &60.0\%  &70.3\% &41.7\% &52.0\%\\
\textsc{Similarity Embedding Network} \cite{huang18aaai}     &     &8  &1.61M   &61.0\%  &78.1\% &45.2\% &65.7\%\\
\textsc{Prototypical Network} \cite{Jake17nips}      &     &3  &1.18M   &67.9\%  &83.0\% &46.2\% &74.2\%\\
\hline
\textsc{Siamese Network} \cite{Koch2015SiameseNN}     &\checkmark     &3+1  &2.50M   &49.3\%  &58.6\% &29.0\% &39.0\%\\
\textsc{Matching Network} \cite{matching_network}    &\checkmark     &3+1  &2.50M   &59.0\%  &74.0\% &38.8\% &55.3\%\\
\textsc{Relation Network} \cite{sung2018learning}    &\checkmark     &7+1  &6.11M   &64.0\%  &74.4\% &46.0\% &57.0\%\\
\textsc{Similarity Embedding Network} \cite{huang18aaai}     &\checkmark     &8+1  &3.40M   &71.2\%  &82.0\% &\textbf{56.9}\% &71.0\%\\
\textsc{Prototypical Network} \cite{Jake17nips}      &\checkmark     &3+1  &2.50M   &\textbf{74.0\%}  &\textbf{87.7\%} &55.0\% &\textbf{76.5\%}\\
\hline

\end{tabular}
\caption{The result of few-shot sound recognition (in \%) on ESC-50. All the baselines reported here are based on our implementation. We indicate whether a method uses attention similarity, the network depth, and the number of model parameters.}
\label{tb:esc50}
\end{table*}

\begin{table*}[t]
\centering
%The \textbf{top}-half compares the performance with several non-attention methods. The \textbf{bottom}-half demonstrates the performance by compared prior art attention networks and the proposed network. We also report the number of depth and parameters of the network.
\begin{tabular}{|l|ccc|cccc|}
\hline
\multirow{2}{*}{Model} & Attentional &\multirow{2}{*}{Depth} &\multirow{2}{*}{Param.}  &\multicolumn{2}{c}{5-way Acc} &\multicolumn{2}{c|}{10-way Acc} \\
\cline{5-8}
  &similarity&&  &1-shot &5-shot &1-shot &5-shot  \\
\hline
\textsc{Siamese Network} \cite{Koch2015SiameseNN}     &     &3  &1.18M   &38.2\%  &43.5\% &25.0\% &28.0\%\\
\textsc{Matching Network} \cite{matching_network}    &     &3  &1.18M   &51.0\%  &61.5\% &31.7\% &43.0\%\\
\textsc{Relation Network} \cite{sung2018learning}     &     &7  &4.78M   &56.2\%  &74.5\% &39.2\% &52.5\%\\
\textsc{Similarity Embedding Network} \cite{huang18aaai}    &     &8  &1.61M   &63.2\%  &78.5\% &44.2\% &62.0\%\\
\textsc{Prototypical Network} \cite{Jake17nips}      &     &3  &1.18M   &66.2\%  &83.0\% &46.5\% &72.2\%\\
\hline
\textsc{Siamese Network} \cite{Koch2015SiameseNN}     &\checkmark     &3+1  &2.50M   &46.0\%  &50.0\% &29.0\% &29.7\%\\
\textsc{Matching Network} \cite{matching_network}    &\checkmark     &3+1  &2.50M   &52.7\%  &66.5\% &36.2\% &48.2\%\\
\textsc{Relation Network} \cite{sung2018learning}    &\checkmark     &7+1  &6.11M   &61.0\%  &76.2\% &40.0\% &59.2\%\\
\textsc{Similarity Embedding Network} \cite{huang18aaai}    &\checkmark     &8+1  &3.40M   &\textbf{70.2\%}  &83.2\% &49.2\% &67.2\%\\
\textsc{Prototypical Network}   \cite{Jake17nips}   &\checkmark     &3+1  &2.50M   &69.7\%  &\textbf{85.7\%} &\textbf{51.5\%} &\textbf{73.5\%}\\
\hline
\end{tabular}
\caption{The result of few-shot sound recognition performance (in \%) on \textit{noise}ESC-50. All our implementation.}
\label{tb:noiseesc50}
\end{table*}

\begin{figure*}[ht]
\centering
\includegraphics[width=.99\textwidth]{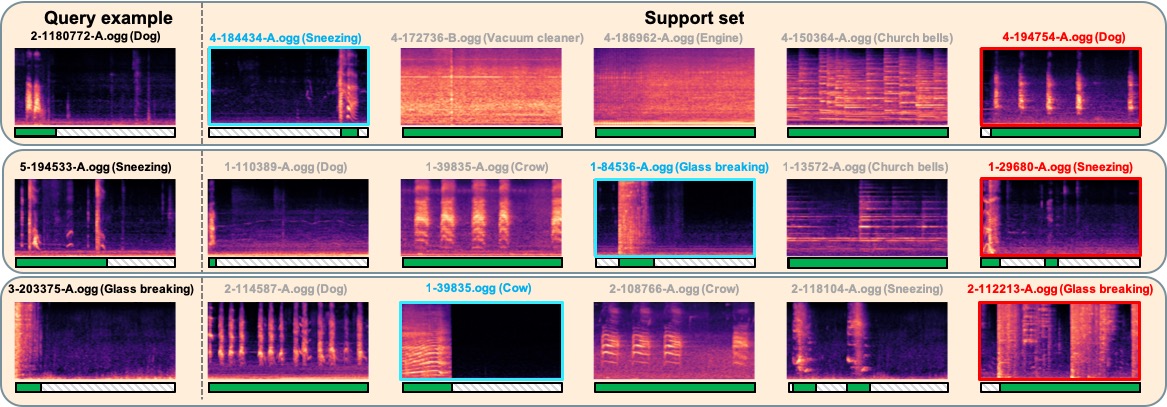}
\caption{Qualitative examples of few-shot sound recognition for 1-shot 5-way task on \textit{noise}ESC-50. Note that the sound events in the test set are never seen during training. The examples marked with red and cyan colors are matched by using the prototypical network with and without attentional similarity, respectively. We manually add the bars in green and gray colors underneath the spectrograms to indicate parts of the audio clips that comprise sounds of interest and the background noise, respectively. }
\label{fig:case_evl} 
\end{figure*}

\subsection{Attentional similarity}

To deal with variable-length inputs, most approaches \cite{huang18aaai,matching_network,oord13nips,Kumar17icassp,Jake17nips} use pooling functions to aggregate the feature maps $\mathbf{X} \in \mathbb{R}^{M \times T}$ to yield a fixed-length vector $\mathbb{R}^{M \times 1}$, where $M$ is the number of channels and $T$ is the number of the temporal dimension. The similarity function can be written as: 
\begin{equation}
f_{sim}(\mathbf{X}_q,\mathbf{X}_j) = dist(pool(\mathbf{X}_q), pool(\mathbf{X}_j)) \,,
\end{equation}
where $pool(\cdot)$ is the pooling function 
%to transform the feature map $\mathbf{X}$ to a vector in $\mathbb{R}^{M \times 1}$.
and $dist(\cdot)$ any distance function between two vectors, such as the inner product.

\textbf{Second-order similarity}: A recent work on second-order similarity estimation \cite{huang18aaai} computes the segment-by-segment (second-order) similarity $\mathbf{X}_q^T\mathbf{X}_j \in \mathbb{R}^{T_q \times T_j}$ between two inputs using the feature maps of the last layer of the ConvNet before pooling. Compared with the fixed-length vector (clip-level feature), this method allows the model to use segment-level feature to learn the temporal correlation between two inputs. The second-order similarity can be written as follows:
\begin{equation}
f_{order2\_sim}(\mathbf{X}_q,\mathbf{X}_j) = \mathbf{X}_q^T\mathbf{X}_j \,.
\label{order2_sim}
\end{equation}
Being inspired by this method, we propose to learn a weight $W_{qj} \in \mathbb{R}^{T_q \times T_j}$ to generate the attentional second-order similarity to capture the importance of segment-by-segment similarity. We can rewrite Eq.~(\ref{order2_sim}) as:
\begin{equation}
f_{order2\_wsim}(\mathbf{X}_q,\mathbf{X}_j) = \mathbf{X}_q^T \mathbf{X}_j W_{qj} \,.
\end{equation}
Following \cite{girdhar17nips}, we can compute attentional similarity by approximating weight $W_{qj}$ as a rank-1 approximation,  $W=\mathbf{A}_j\mathbf{A}_q^T$ where $\mathbf{A}_j,\mathbf{A}_q \in \mathbb{R}^{T_j \times 1},\mathbb{R}^{T_q \times 1}$. Then, we can derive the following \emph{attentional similarity} function:
\begin{align}
f_{att\_sim}(\mathbf{X}_q,\mathbf{X}_j) =& Tr(\mathbf{X}_q^T \mathbf{X}_j\mathbf{A}_j\mathbf{A}_q^T)\,,\\
=& Tr(\mathbf{A}_q^T\mathbf{X}_q^T \mathbf{X}_j\mathbf{A}_j) \,,\\
=& \mathbf{A}_q^T(\mathbf{X}_q^T \mathbf{X}_j)\mathbf{A}_j\,,
\label{eq:attsim}
\end{align}
where $Tr(\cdot)$ is the trace operator and $\mathbf{A}_q$ is the attention vector computed by using another stack of convolutional layers $f_{att}(\cdot)$ by feeding $\mathbf{X}_q$ to find the important of segments. Eq.~(\ref{eq:attsim}) uses the attention vector $\mathbf{A}_q$ and $\mathbf{A}_j$ to compute a weighted average of segment-by-segment similarity. More importantly, Eq.~(\ref{eq:attsim}) can be rewritten as follows:
\begin{align}
f_{att\_sim}(\mathbf{X}_q,\mathbf{X}_j) =& \mathbf{A}_q^T(\mathbf{X}_q^T \mathbf{X}_j)\mathbf{A}_j\,,\\
=&(\mathbf{X}_q \mathbf{A}_q)^T (\mathbf{X}_j\mathbf{A}_j) \,.
\label{eq:fattsim}
\end{align}
The final equation can be interpreted as we compute the similarity score by using the inner product between two attentional vector $\mathbf{X}_q \mathbf{A}_q$ and $\mathbf{X}_j \mathbf{A}_j$. This allows us to replace the inner product with common distance functions (e.g., cosine similarity or Euclidean distance) to measure the distance between two attentional vectors. 
%So, the attentional similarity can be succinctly written as $(\mathbf{X}_q \mathbf{A}_q)^T (\mathbf{X}_j\mathbf{A}_j)$.
So, in general the attentional similarity can also be computed as $dist(\mathbf{X}_q \mathbf{A}_q, \mathbf{X}_j\mathbf{A}_j)$.

\section{Experiments}
\label{sec:experiments}

\subsection{Experimental Settings}
\textbf{Dataset}: We conduct experiments on few-shot sound recognition using two datasets: ESC-50 and \textit{noise}ESC-50. The ESC-50 dataset \cite{piczak2015dataset} contains 2,000 5-seconds audio clips labeled with 50 classes, each having 40 examples. The sound categories cover sounds of \texttt{Animals}, \texttt{Natural}, \texttt{Human} and \texttt{Ambient noises}. To evaluate the models under background noise conditions, following \cite{DCASE17}, we create the second dataset, coined \textit{noise}ESC-50, by augmenting ESC-50 audio clips with additive background noise randomly selected  from audio recordings of 15 different acoustic scenes from the DCASE2016 dataset \cite{Mesaros2016TUTDF}. Such a synthetic strategy allows to generate artificially noisy audio examples that reflect the sound recordings in everyday environment. Therefore, evaluation on \textit{noise}ESC-50 may better measure how the models perform in real-world applications. 

We note that, although the size and vocabulary of ESC-50 is small and limited, it is appropriate to use it as a public benchmark dataset for few-shot sound recognition. Larger benchmark datasets such as AudioSet \cite{audioset} may suffer from the openness issue of audio data \cite{Fonseca2017freesound} and class imbalance problems \cite{huang2016lmle}.

\textbf{Data preparation}: In order to directly compare our model against strong baselines for few-shot learning, our experiment uses the similar splits proposed by \cite{matching_network}. The 50 sound event classes from ESC-50 dataset are divided into 35 classes for training and 10 classes for test. We train our model and baselines on 35 classes and use the remaining 5 validation classes for selecting the final model.

\textbf{Feature extraction}: To speed up model training, all audio clips from the ESC-50 and DCASE2016 datasets are downsampled from 44.1 kHz to 16 kHz.
We extract the 128-bin log mel-spectrogram from raw audio as the input feature to the neural networks. The \texttt{librosa} library \cite{mcfee15scipy} is used for feature extraction. Before training a model, the input features are z-score normalized using the mean and standard deviation coefficients computed from the training set.

\textbf{Network design}: The backbone network is based on the simple yet powerful CNN structure, which has been widely used in audio tasks \cite{oord13nips,Kumar17icassp}. The input feature of network is a mel-spectrograms, with 128 frequency bins and 160 frames. Our backbone network consists of a stack of blocks, each of which has a $3\times3$ convolutional layer followed by batch normalization \cite{Sergey15icml}, a ReLU activation layer and a $4\times4$ max-pooling layer. For optimization, we use stochastic gradient descent (SGD) and initial learning rate of 0.01. The learning rate is divided by 10 every 20 epochs for annealing, and we set the maximal number of epochs to 60. Moreover, we set the weight decay to 1e-4 to avoid overfitting. 

Please note that all the baselines and our model are used the same backbone network. In our pilot studies, we have also explored using other advanced CNN structure such as ResNet \cite{He16cvpr} or VGG \cite{Simonyan15} as our backbone network but seen no much improvement. This may be due to the moderate size of the ESC-50 dataset. Code are available at \url{https://github.com/kevinco27/attentional-similarity}.

%\vspace{-3mm}
\subsection{Experimental Results}
Table \ref{tb:esc50} compares the performance of our own implementation of five metric-based learning methods for few-shot sound recognition with and without the proposed attentional similarity module on ESC-50. We can see that using attentional similarity clearly improves all the existing methods, giving rise to +4\% to +7.1\% relative improvement in 5-way 5-shot learning, a large performance gain. According to our experiment, the prototypical network turns out to be more effective and efficient than the relation network and the similarity embedding network for few-shot sound recognition. This seems to support similar findings in computer vision tasks \cite{Jake17nips,sung2018learning}. 

Table \ref{tb:noiseesc50} shows the experimental result on \textit{noise}ESC-50. Again, we see that the attentional similarity module consistently improves the result of existing methods, and the relative performance gain ranges from +2.1\% to +6.5\% in 5-way 5-shot learning. Compared with ESC-50, the performance gain decreases slightly, possibly because sound event recognition over \textit{noise}ESC-50 is more challenging. Finally, similar to the case in ESC-50, the attentional similarity-empowered prototypical network achieves the best result among the evaluated methods by a great margin in 5-shot learning.

Figure \ref{fig:case_evl} gives a qualitative comparison of the result of the 
prototypical network with and without attentional similarity for 3 query examples from the \textit{noise}ESC-50 test set. Those picked by the model without attentional similarity (i.e., marked in cyan) do not share the same class as the queries; they are picked possibly because both the query and the picked one have a long silence. In contrast, the model with attentional similarity finds correct matches (marked in red).
%All examples suggest that our attentional similarity can boost the performance of few-shot sound recognition by matching relative shot or transient sound events.  

\section{Conclusion}
\label{sec:concl&future work}
We have introduced a simple module of attentional similarity for few-shot sound recognition to generate an attentional representation of inputs. It allows the model to ignore the unrelated background noise while matching relative short sound events. Extensive experiments show that attentional similarity consistently improves the performance of various existing methods on datasets of either noise-free or noisy clips.  
In the future, we plan to extend the model to adopt a multi-label learning setting for few-shot sound recognition.
%Generally, sound recognition is a multi-label classification task. 
%For future work, we are interested in combining the task of multi-label learning with few-shot sound recognition.  

% Below is an example of how to insert images. Delete the ``\vspace'' line,
% uncomment the preceding line ``\centerline...'' and replace ``imageX.ps''
% with a suitable PostScript file name.
% -------------------------------------------------------------------------

%\vfill
%\pagebreak

%\vfill\pagebreak

\bibliographystyle{IEEEbib}
\bibliography{refs}
\end{document}